\documentclass[9pt]{article}

\usepackage{authblk,epsfig,graphics}
\usepackage{float}
\usepackage{amssymb,amsmath}
\usepackage{pdfpages}
\usepackage{enumitem}

\newcommand{\be}{\begin{equation}}
\newcommand{\ee}{\end{equation}}

\title{Numerical analysis of scattering on combinatorial graphs}

\author[1]{\normalsize{Moysey Brio}\thanks{brio@math.arizona.edu}}
\author[2]{\normalsize{Jean-Guy Caputo }\thanks{jean-guy.caputo@insa-rouen.fr}}

\affil[1]{Department of Mathematics, University of Arizona,
617 North Santa Rita Avenue, Tucson Arizona, 85721, United States of America}
\affil[2]{Laboratoire de Math\'ematiques, INSA de Rouen Normandie,
Avenue de l'Universit\'e, Saint-Etienne du Rouvray, 76801, France}

\date{\ }

\begin{document}

\maketitle

\begin{abstract}
We investigate numerically the scattering of waves on discrete graphs. An efficient algorithm is developed to compute the reflection and transmission (spectral) coefficients. We then explore various configurations of input and output leads, demonstrating how bound states—arising from specific vertex-lead connections—result in total reflection. The impedance of the leads is shown to influence the spectral coefficients in a predictable manner. Furthermore, for a given input lead we show that the total transmission of a wavepacket can be maximized by appropriately selecting the exit lead. Finally, we analyze the spectral signatures of defects within the graph and find that they vary depending on both the defect's location and its spectral characteristics thus enabling its identification.
\end{abstract}


\tableofcontents 

\section{Introduction}

Building on Linus Pauling’s foundational contribution \cite{pauling}, quantum graphs--together with their spectral and scattering properties--have become a versatile and effective framework for modeling and analyzing wave propagation and vibrational dynamics in a wide variety of complex, quasi-one-dimensional quantum systems in both physics and chemistry. In physics, quantum graphs and
metric graphs have been applied to the study of quantum chaos \cite{gnutzmann}, design and analysis of metamaterials \cite{lawrie}, waveguide dynamics in fiber optics and superconducting networks \cite{dietz17,dietz24}. In chemistry, they have been used to model vibrational dynamics in molecules \cite{fabry} and to explore protein structure \cite{gadi}. Quantum graphs can be simulated using networks based on superconducting circuits, coaxial microwave cables, and optical fiber waveguides \cite{dietz24, akhshani,ahmed}. A central challenge in the development of novel quantum devices lies in the precise control and manipulation of their scattering and transport properties, as well as ensuring their robustness and operational efficiency \cite{ahmed,chen,yusupov}.

Discrete or combinatorial graphs can be seen as approximations of
quantum or metric graphs. For these systems, we only consider
functions defined at the vertices of the graph. The graph Laplacian
matrix i.e. the degree matrix minus the adjacency matrix captures
the connectivity of the graph, see \cite{chung,crs01} for precise
definitions. In contrast, the 
quantum graph Laplacian is a differential operator defined on functions 
over the edges of a metric graph, with connectivity encoded via boundary 
conditions at the vertices. When edge lengths of the metric graph shrink 
to zero, the spectrum and eigenfunctions of the quantum graph Laplacian 
can, under certain limiting conditions, be approximated by those of a 
graph Laplacian \cite{post09,post06}. 
Graph Laplacians and adjacency matrices have been applied to many areas,
including localized oscillations in enzymes \cite{farias}, mass-spring models \cite{piazza}, the electrical grid \cite{ckr19}, fluid networks \cite{todini}, machine learning \cite{zhu,belkin04}, neural networks \cite{kipf}, and data science \cite{belkin03}.

In \cite{drinko19,drinko20,afshin}, the scattering properties of 
continuous quantum graphs with incoming and outgoing leads were 
analyzed to investigate the transmission amplitudes of 
simple quantum graphs composed of regular polygons, such as 
triangles and squares. These studies revealed the existence of a 
forbidden (transmission) band. The authors also used the
internal configuration of the graphs to control wave packets.
The scattering properties of discrete or combinatorial
graphs have mainly focused on the adjacency matrix.
Colin de Verdiere and Truc \cite{colin13} established a scattering 
theory for graphs isomorphic to a regular tree at infinity, see
also the study by Fedorov and Pavlov \cite{fedorov06}.
The formalism was used by Fowler et al \cite{fpt14} 
to select molecules with a large electrical conductivity. 
For the graph Laplacian, a pioneering study by Smilansky \cite{smilansky}
established the foundations of scattering theory.
In the context of inverse problems, Novikov~\cite{novikov} studied the problem of source detection on an infinite grid for the discrete Helmholtz operator, employing scattering theory. Most of these investigations are theoretical in nature, which gives rise to several practical questions, including the following:
\begin{itemize}[leftmargin=10pt]
    \item What is a practical algorithm to compute the transmission and reflection coefficients for a given graph?
    \item How do these coefficients change when the vertices connected to the leads are modified?
    \item Which lead maximizes the transmission of a wave packet through the graph?
    \item How does the impedance of the leads influence scattering?
    \item Is it possible to detect a source emitting on a graph through a scattering experiment?
\end{itemize}

To address these questions, in this article, we begin by presenting a motivating example involving a scattering problem on a triangular graph with two attached leads. We then provide a review of general scattering theory, including explicit expressions for the reflection and transmission coefficients with special attention given to the phenomena of resonances, as well as conditions for total transmission and total reflection. Subsequently, we investigate the role of lead-graph coupling—interpreted as the "impedance" at the interface—through several illustrative configurations. Using a set of simple graphs with two or three attached leads, we demonstrate various theoretically possible scenarios of total transmission and total reflection arising from the graph’s bound states. We further analyze how the scattering properties depend on the impedance strength, exploring how transmission can be optimized, suppressed, or switched based on the specific vertices to which the leads are connected. Finally, we address the inverse problem of source detection and localization, examining how the signature of a source
depends on its location within the graph and on its frequency.

The structure of the article is as follows. Section 2 outlines the problem statement and provides a concise review of the relevant background on the scattering problem for the discrete wave equation on graphs with attached leads, including explicit expressions for the reflection and transmission coefficients. In Section 3, we investigate the influence of bound states of the graph Laplacian, the role of vertex connectivity to the leads, and the impact of interface impedance on the scattering behavior. Additionally, we address strategies for optimizing the transmitted energy. Section 4 presents a case study on source detection, where the source is positioned at various graph vertices and emits at a fixed frequency. Finally, Section 5 offers concluding remarks on the findings and discusses potential directions for future research.

\section{The graph wave equation and scattering formalism}

\begin{figure} [H]
\centerline{
   \epsfig{file=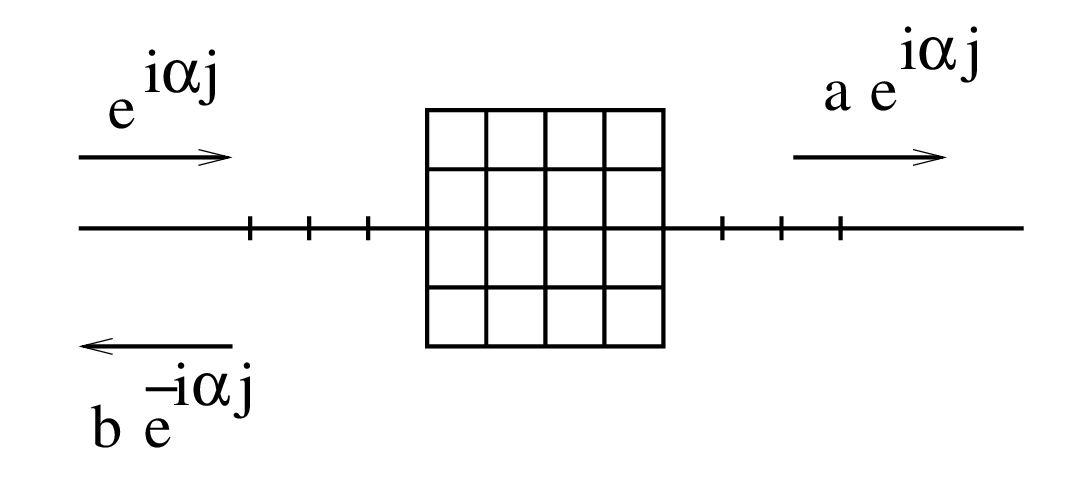,height=4 cm,width=12 cm,angle=0}
}
\caption{
Two chains connected to a grid graph.
}
\label{schema}
\end{figure}

We consider a graph $G$ with $n$ vertices to which are connected
two or more infinite chains, see Fig. \ref{schema}. In the following, 
to connect with practical wave applications, we refer to these chains as leads.
The wave dynamics of miscible fluids on the graph is governed 
by the following wave equation \cite{cks13}
\be \label{gwave}  {\ddot u} - L u = F(x,t) ,\ee
where $u=(u_1,u_2,\dots)$, $L$ is the Laplacian matrix of the graph 
and $F(x,t)$ a vector
source term that can be time dependant. 
Since the problem is linear, it is natural to separate time and space and
write
\be \label{harm} u(t)= e^{i k t} \psi .\ee
This yields the following Helmoltz problem
\be \label{ghelm}  L \psi  + k^2 \psi=  F_k,\ee
where $F_k$ is the $k$ Fourier component of $F$.

We begin by assuming the absence of a source term, i.e.,  $F=0$. In a recent study \cite{gcckp20}, we provided a complete characterization of the bound states—namely, the eigenvalues and eigenvectors of the Laplacian—associated with a complete graph coupled to infinite chains. In the present work, we focus on the traveling waves component of the scattering data. To this end, we consider the equation governing free wave propagation along an arbitrary lead away from the graph vertices.

On each lead extending from the central graph $G$ the solution to the one-dimensional chain Laplacian satisfies the following equation, with the lead index $l$ suppressed for brevity.
\be
-2 \psi_j + \psi_{j-1}+\psi_{j+1} + k^2 \psi_j=0, \;\;j\ge 0.   
\label{1dLaplacian}
\ee
Equation \eqref{1dLaplacian} has an exact propagating solution of the form 
$$\psi_j=e^{i \alpha(k) j}, $$ where we restrict frequencies $k$ to 
be nonnegative, $k\ge 0$. After substitution we get 
\be \label{eqn:char}
-2  + e^{-i \alpha(k)}+  e^{i \alpha(k)} + k^2 =0, 
\ee
or 
\be
( e^{-i \alpha(k)/2}-  e^{i \alpha(k)/2})^2 + k^2 =0, 
\ee
that simplifies to 
$$\displaystyle {\sin^2\big(\frac {\alpha(k)}{2}\big)}=\frac{k^2}{4} ,$$ 
resulting in the following dispersion relation for propagating solutions, 
\be \label{disChain}
\alpha(k)=2 \arcsin\big(\frac{k}{2}\big), \;\;  \text{for}\;\; 0\le k \le 2\;\;  \text{that corresponds to}\;\; 0\le \alpha(k) \le \pi.
\ee

We now consider the complete problem, which consists of two semi-infinite chain graphs attached to a central graph~$G$. A wave is incident from one of the chains, and we analyze the resulting reflected and transmitted components. This setup constitutes a \emph{scattering problem}. To illustrate the scattering process, we present a detailed calculation for a specific example below.

\subsection{Example of Unidirectional Scattering} 

\begin{figure} [H]
\centerline{
   \epsfig{file=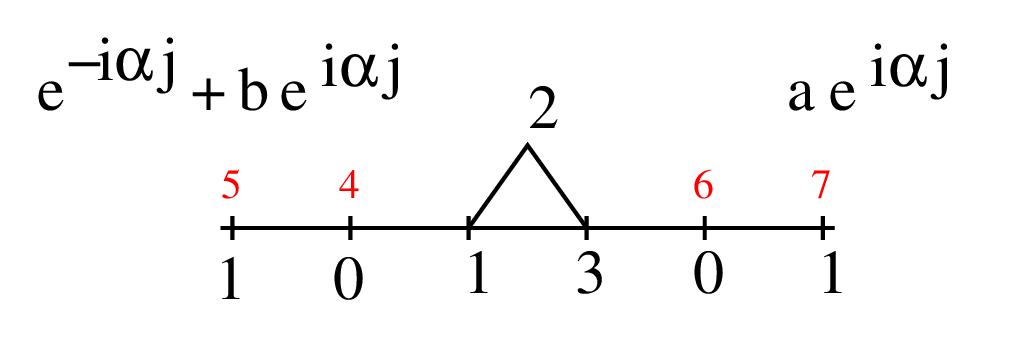,height=4 cm,width=12 cm,angle=0}
}
\caption{
Two chains connected to a complete graph $K_3$. The incident and 
reflected waves (shown on the left)
 and the transmitted waves (shown on the right) from the chains are indicated.
}
\label{k3}
\end{figure}

Consider the example where the central graph
is the complete graph $K_3$ with 3 vertices, see  Fig. \ref{k3}.
Writing the Helmholtz problem (\ref{ghelm}), we obtain
\begin{align*}
-3 \psi_1 + \psi_4 + \psi_2 + \psi_3  + k^2 \psi_1 &= 0 ,\\
-2 \psi_2 + \psi_1 + \psi_3 + k^2 \psi_2 &= 0 ,\\
-3 \psi_3 + \psi_1 + \psi_2 + \psi_6 + k^2 \psi_3 &= 0 ,\\
-2 \psi_4 + \psi_1 + \psi_5 + k^2 \psi_4 &= 0 ,\\
-2 \psi_6 + \psi_3 + \psi_7 + k^2 \psi_6 &= 0 .
\end{align*}
Assuming an incident and reflected wave on the left chain and a transmitted wave on the right chain,
 then $\psi_4, \psi_5, \psi_6$ and $\psi_7$ are specified as follows:

\begin{align*}
\psi_4 &=1 + b ,\\
\psi_5 &= e^{-i\alpha} + b(k) e^{i\alpha} ,\\
\psi_6 &= a(k) ,\\
\psi_7 &= a(k) e^{i\alpha} ,
\end{align*}
where the dependence on $k$ for $\alpha(k)$ was omitted for brevity.
Note that the chains are oriented outwards from G and parameterized with $0, 1, \dots $  as shown in Fig.\ref{k3}.

\begin{figure} [H]
\centerline{
   \epsfig{file=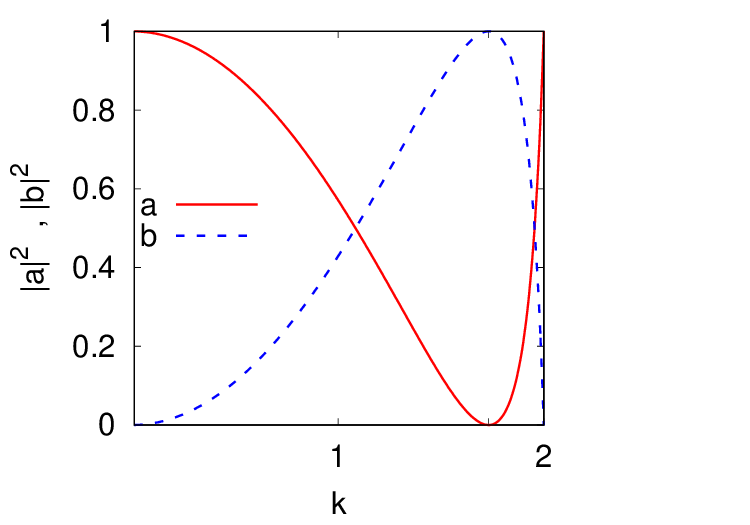,height=4 cm,width=12 cm,angle=0}
}
\caption{ Plot of $|a|^2(k)$ and $|b|^2(k)$ for the 
graph of Fig. \ref{k3}.
}
\label{abk3}
\end{figure}

Finally, using the dispersion relation, we obtain the linear system
\begin{align} \label{sysEX}
\begin{pmatrix}
k^2-3   & 1     & 1     & 0 & 1\\
1       & k^2-2 & 1     & 0 & 0\\
1       &   1   & k^2-3 & 1 & 0\\
0       &  0    & 1     & -e^{-i\alpha} & 0\\
1       &  0     & 0    & 0 & -e^{-i\alpha}
\end{pmatrix}
\begin{pmatrix}
\psi_1 \\
\psi_2 \\
\psi_3 \\
a \\
b
\end{pmatrix}
= 
\begin{pmatrix}
- 1 \\
0 \\ 
0 \\ 
0 \\ 
e^{i\alpha} \\
\end{pmatrix},
\end{align}
which can be solved in a standard way.
Then one can sweep a range of values of $k$ to get
$a(k), b(k)$. The values $|a|^2(k)$ and $|b|^2(k)$ are reported in Fig. \ref{abk3}.
We can make the following remarks
\begin{itemize}[leftmargin=10pt]
\item As expected, we have $$ |a|^2 + |b|^2=1  .$$
\item We have total reflection $|b|=1, |a|=0$ for $k=\sqrt{3}$, 
see tick on the x axis in Fig. \ref{abk3}.
This value of $k$ corresponds to the eigenvalues of the 
graph Laplacian for the graph $K_3$
$$\lambda=0, 3^2,$$ 
where the exponent corresponds to the multiplicity of the eigenvalue. As expected 
the transmission is 0 when we excite a bound state. We justify this below. 
\item The graph becomes transparent, i.e., $|a|=1$, for $k=0$ and $2$. 
\end{itemize}

\subsection{Scattering formalism }

Following the approach in~\cite{smilansky}, we consider the graph Laplacian 
associated with a graph $G$, to which semi-infinite leads are attached at 
selected nodes of $G$. Let $\vec{\phi}$ denote the wavefunction values at the nodes belonging to the graph $G$, and let $\vec{\psi}_j^l$ represent the wavefunction values at the nodes on the semi-infinite leads. Here, the index $l$ labels the individual leads, while the index $j$ indicates the position along each lead.
Each lead is oriented in the positive direction, extending away from the graph. The components with $j = 0$ correspond to the interface nodes, where each lead is connected to the graph. The components with $j = 1, 2, 3, \dots$ represent nodes further along the lead, progressively farther from the graph.

The vectors $\vec{\phi}$ and $\vec{\psi}$ satisfy a coupled system of linear equations: one set governs the dynamics of the wavefunction $\vec{\phi}$ on the graph $G$, and the other governs the wavefunction on the semi-infinite leads for the interface values $\vec{\psi}_0$ that mediate the coupling between the graph and the leads.
The assumption of continuity at the boundary nodes between 
the lead and the graph implies that $\vec \psi_{-1}$ equals
the corresponding values of the graph nodes. 
 Consequently, on each lead the free propagating solution 
at the nodes $j=0,1$ can be written as a superposition of two 
counter-propagating waves satisfying the dispersion relation (\ref{disChain}), 
the incident and reflected waves (also often called incoming and 
outgoing waves), as 
\be \label{1dwaves}
\begin{split}
& \vec \psi_0=\vec a +\vec b, \\ 
& \vec \psi_1=\vec a e^{-i \alpha(k)} +\vec b e^{i \alpha(k)},  
\end{split}
\ee
where vectors $\vec a=[a_l]$ and $\vec b=[b_l]$ contain respectively
the amplitudes of the incident and reflected
waves for the leads $l=1,2,3,\dots, \ell$.   
In our further discussion we will encounter two boundary conditions, 
$\vec \psi_0=0$ (Dirichlet) and $\vec \psi_0=\vec \psi_{1}$ (Neumann). Both 
boundary conditions result in perfect reflections with different phase shifts, 
\begin{align}
& \vec b =-\vec a=e^{-i\pi} \vec a,\;\; \text{(Dirichlet),} \\ 
& \vec b =\frac{ e^{i \alpha(k)}-1}{1-e^{-i \alpha(k)}} \;\vec a=e^{i \alpha(k)} \vec a,\;\; \text{(Neumann).}   
\end{align}

We next consider the graph Laplacian associated with the combined structure 
consisting of the graph $G$ and its attached leads. The full Laplacian is 
defined as $\displaystyle L=-D +A$, 
 
where $D$ is the diagonal degree matrix whose entries correspond to the degree of node $i$, that is, the total number of nodes connected to node $i$, including connections to adjacent lead nodes at position $j=0$.The matrix $A=[a_{ij}]$ is the adjacency matrix, with $a_{i,j}=1$ if nodes $i$ and $j$ are connected, and $a_{i,j}=0$
otherwise.

For simplicity, we assume that the graph contains no loops or multiple edges (i.e., no parallel bonds), and that at most one lead is attached to any given node of $G$. It is useful to decompose the Laplacian 
$L$ into contributions from the graph and the leads. Specifically, we write
where:
 $\displaystyle L=-D+A=-D^0-\tilde D +A^0+ W = L^0-\tilde D+W$, 
where the diagonal matrix $\tilde D$ has diagonal value set to 1 
for the graph nodes with the leads attached and 0 for the nodes without 
leads. The matrix $A^0$ indicates the inner connectivity of the graph 
nodes $\vec \phi$, and $L_0$ is the graph Laplacian for the graph $G$
alone excluding the leads. 

Let $W \in \mathbb{R}^{n \times \ell}$ denote the \emph{lead matrix}, where $n$ is the number of nodes in the graph and $\ell$ is the total number of leads. The entries of $W$ are defined as
\[
w_{ij} = 
\begin{cases}
1, & \text{if node } i \text{ is connected to lead } j, \\
0, & \text{otherwise}.
\end{cases}
\]
Note that the transpose $W^T$ encodes the connections from leads to graph nodes—that is, it represents the connectivity structure from the perspective of the leads. Consequently, the following matrix identities hold
\[
W W^T = \tilde{D}, \quad W^T W = I_m,
\]
where $\tilde{D}$ is a diagonal matrix and $I_m$ is the identity matrix of size $m$.

The equation for the graph nodes can be written as 
\be 
(L_0-\tilde D +k^2 I_n)\vec \phi +  W \vec \psi_0=0, \label{eqn:inner} 
\ee
and the equation for the lead nodes adjecent to the graph, $\vec \psi_0,$ as 
\be
-2 \vec \psi_0 +\vec \psi_1 + W^T \vec \phi +k^2 \vec \psi_0=0.  \label{eqn:leads} 
\ee

Finally, we slightly generalize the system of equations by allowing the 
strength of the coupling between the graph $G$ and the attached leads -the so-called impedance- to be different from 1 and replace $L_0$ 
with $v^2 L_0$ in the equation \eqref{eqn:inner}. Note that $v$ may also 
be interpreted as the speed of propagation of interactions within the graph.

To summarize, in order to determine the values at the graph nodes \( \vec{\phi} \) and the lead nodes \( \vec{\psi} \), the following coupled system of equations must be solved:

\begin{align} 
 (v^2 L_0-\tilde D +k^2 I_n)\; \vec \phi +  W \vec \psi_0 &=0,\label{eqn:system1}  \\ 
 -2 \vec \psi_0 +\vec \psi_1 + W^T \vec \phi +k^2 \vec \psi_0& =0.\label{eqn:system2}  
\end{align}
This system can be solved by substitution. Use equation \eqref{eqn:system1}  to express 
$G$ nodes $\vec \phi$ in terms of the adjacent lead values $\vec \psi_0$,    
\be
\vec \phi= F^{-1} W \vec \psi_0,
\ee
where the matrix 
$$F \equiv -(v^2 L_0-\tilde D +k^2 I_n) = -(v^2 L_0-W W^T +k^2 I_n) $$ 
is at the moment assumed to be invertible (non-resonant case). Substituting 
into equation \eqref{eqn:system2} gives 
\be
-2 \vec \psi_0 +\vec \psi_1 + W^T F^{-1} W \; \vec \psi_0 +k^2 \vec \psi_0=0, 
\ee
that can be written in terms of the incident and reflected amplitudes using equations \eqref{1dwaves} as,
\be
-2 (\vec a +\vec b) + \vec a e^{-i \alpha(k)} +\vec b e^{i \alpha(k)} + W^T F^{-1} W \; \vec (\vec a +\vec b  ) +k^2 (\vec a  +\vec b  )=0. 
\ee
Now, applying the dispersion relation \eqref{eqn:char}, \;  $\displaystyle 2-k^2= e^{-i \alpha(k)}+  e^{i \alpha(k)},$ and solving for the reflected wave amplitudes in terms of the incident wave amplitudes gives the scattering equation, 
\be \label{scatEQ}
 \vec b=S \vec a,  \ee
where the scattering matrix $S$ is  
\be
\displaystyle S=-(I_m \;  e^{-i \alpha(k)} -W^T F^{-1} W )^{-1} \; (I_m \; e^{i \alpha(k)} -W^T F^{-1} W).
\ee
Since the matrix $\displaystyle W^T F^{-1} W $ is real symmetric, the scattering matrix is singular 
only for cases when $\displaystyle e^{-i \alpha(k)}$ is real.  For $0\le \alpha(k)\le \pi$ is 
real only for at the endpoints of the propagating spectrum, i.e., at \( k = 0 \) and \( k = 2 \).

In practical computations of reflection and transmission, we assume the incident wave vector \( \vec{a} \) is given and solve for \( \vec{\phi} \) and \( \vec{b} \). The system of equations \eqref{eqn:system1}--\eqref{eqn:system2} can then be written as a single block matrix equation: 
\begin{align}\label{scatSys}
\begin{pmatrix}
v^2 L_0- W W^T + k^2 I_n   & W \\
W^T                        & -e^{-i \alpha} I_\ell
\end{pmatrix}
\begin{pmatrix}
\vec \phi \\
\vec b 
\end{pmatrix}
= 
\begin{pmatrix}
-W \vec a\\
e^{i \alpha} \vec a 
\end{pmatrix}
\end{align}
This system coincides with (\ref{sysEX}) if $G=K_3$.
 Specifically, for \( n = 3 \), \( \ell = 2 \), the matrices and vectors are given by:

\[
L_0 =
\begin{pmatrix}
-2 & 1 & 1 \\
1 & -2 & 1 \\
1 & 1 & -2
\end{pmatrix}, \quad
W =
\begin{pmatrix}
0 & 1 \\
0 & 0 \\
1 & 0
\end{pmatrix}, \quad
\vec{a} =
\begin{pmatrix}
0 \\
1
\end{pmatrix}, \quad
\vec{b} =
\begin{pmatrix}
a \\
b
\end{pmatrix}.
\]

In the propagating range \( 0 < \alpha(k) < \pi \), the scattering matrix \( S \) is nonsingular. The matrices
\[
H = \operatorname{Im} e^{-i\alpha(k)} - W^\top F^{-1} W, \quad
H^* = \operatorname{Im} e^{i\alpha(k)} - W^\top F^{-1} W
\]
are complex conjugates of each other and commute. This implies that
\begin{align}
S S^* &= H^{-1} H^* (H^{-1} H^*)^* = H^{-1} H^* H (H^*)^{-1} \nonumber \\
&= H^{-1} H H^* (H^*)^{-1} = I_\ell, \label{eq:unitarity}
\end{align}
and similarly \( S^* S = I_\ell \). Therefore, \( S \) is unitary, and it follows that \( \| \vec{a} \|^2 = \| \vec{b} \|^2 \), as expected from th energy conservation.

\vspace{5pt}
\noindent
{\bf There are two notable special cases for the linear system~\eqref{scatSys}:}

\begin{itemize}[leftmargin=10pt]
\item \textbf{Graph resonance:}  
If the matrix \( F = -\left( v^2 L_0 - \widetilde{D} + k^2 I_n \right) \) is singular, then there exists a nontrivial solution to
\[
\left( v^2 L_0 - \widetilde{D} + k^2 I_n \right) \vec{\phi} = 0
\]
for some \( k \in (0, 2) \). In this case, equation~\eqref{eqn:system1} implies \( \vec{\psi}_0 = 0 \), corresponding to perfect reflection under Dirichlet boundary conditions on a semi-infinite lead, as discussed previously.

\item \textbf{Neumann boundary condition and internal graph resonance:}  
If \( k \) is an eigenvalue of the internal graph Laplacian and \( \vec{\phi} \) is the corresponding eigenvector, then
\[
\left( v^2 L_0 + k^2 I_n \right) \vec{\phi} = 0,
\]
resulting in perfect reflection under Neumann boundary conditions. For example, this occurs when \( k = \sqrt{3} \) in the case of the triangle graph from equation (2), where \( \vec{a} = \phi_3 = 0 \) and \( |b| = 1 \).
\end{itemize}

At the endpoints of the $k$ interval, namely at $k = 0$ and $k = 2$, the characteristic roots \( e^{\pm i \alpha(k) j} \) coincide and are equal to \( \pm 1 \), respectively. The corresponding general solution along each lead (omitting the superscript $l$) is given by
\[
\psi_j = a \, (\pm 1)^j + b \, j \, (\pm 1)^j.
\]
Solutions with \( b \ne 0 \) are unphysical, as they exhibit spatial exponential growth away from the graph along the leads. Therefore, the only physically acceptable solution on each lead is \( \psi_j = (\pm 1)^j \), respectively.

The associated linear system for \( \vec{\phi} \), with \( \psi_0 = 1 \) on each lead, is given by equation~\eqref{eqn:system1}. For \( k = 0 \), this system has at least one solution in which all node values are equal to one. 

For \( k = 2 \), the existence of a solution to equation~\eqref{eqn:system1} depends on whether \( -k^2 = -4 \) lies in the spectrum of the matrix \( v^2 L_0 - \widetilde{D} \). In the former case, when
\[
\left( v^2 L_0 - \widetilde{D} + k^2 I_n \right) \vec{\phi} = 0,
\]
the system~\eqref{eqn:system1} is incompatible, since \( \vec{\psi}_j \) is nonzero. In the latter case, when the matrix \( F \) is invertible, we may solve equation~\eqref{eqn:system1} with \( \vec{\phi} = 0 \).


\section{Numerical results}

In this section, we compute the transmission and reflection coefficients for various configurations.
In particular, we investigate cases where more than two leads are connected to the graph. 
We begin by analyzing how the specific vertices at which these leads are attached influence the spectral coefficients.
Next, we explore the impact of the impedance parameter $v$ on the scattering process. 
Finally, we examine how to optimize transmission through the graph as a function of the lead attachment points and the frequency domain.

\subsection{Influence of the connection vertex: relation with eigenvectors}

\begin{figure} [H]
\centerline{
   \epsfig{file=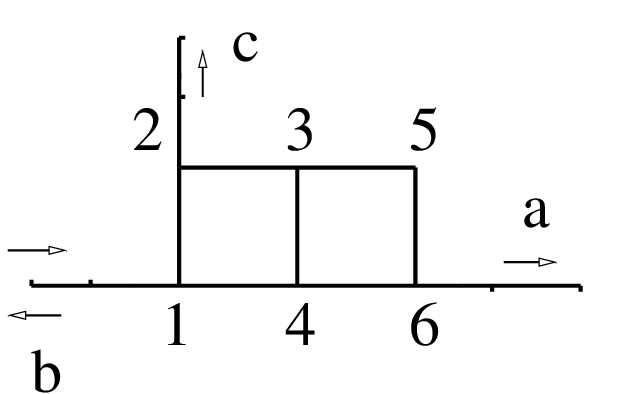,height=4 cm,width=12 cm,angle=0}
}
\caption{
A graph exhibiting broken symmetry between the left-right and 
left-top branches. The amplitudes $a,b,c$ of the spectral coefficients
are indicated.
}
\label{gsqaa}
\end{figure}

We now consider a scenario where three chains are connected to a central graph and 
analyze 
how a wave entering through one of the chains is scattered by the graph. 
The most interesting case arises when symmetry is broken between the
 top and right branches. Consider the graph shown in Fig.~\ref{gsqaa}.

\begin{figure} [H]
\centerline{
   \epsfig{file=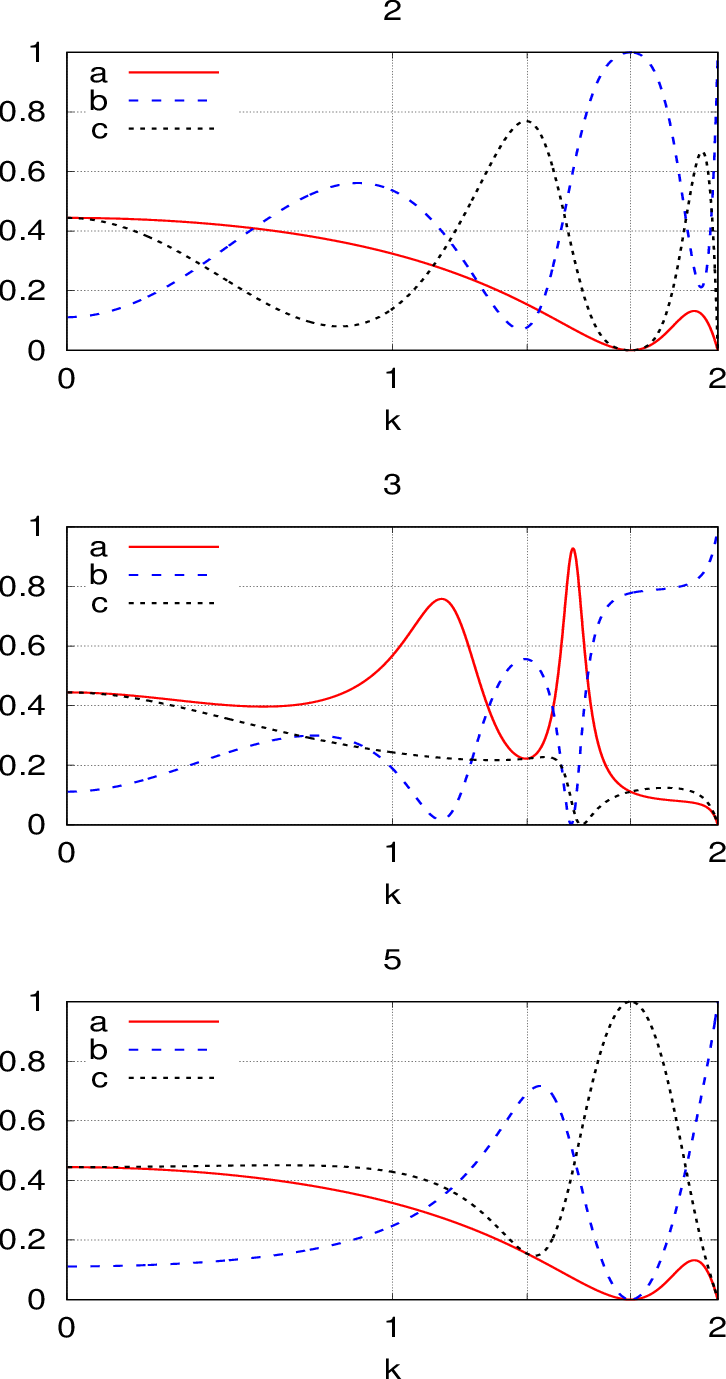,height=8 cm,width=12 cm,angle=0}
}
\caption{
Plots of $|a|^2(k)$, $|b|^2(k)$, and $|c|^2(k)$ for the graph shown above,
with the top branch connected to vertex 2 (top panel), vertex 3 (middle panel),
and vertex 5 (bottom panel).
}
\label{con235}
\end{figure}

\begin{figure} [H]
\centerline{
   \epsfig{file=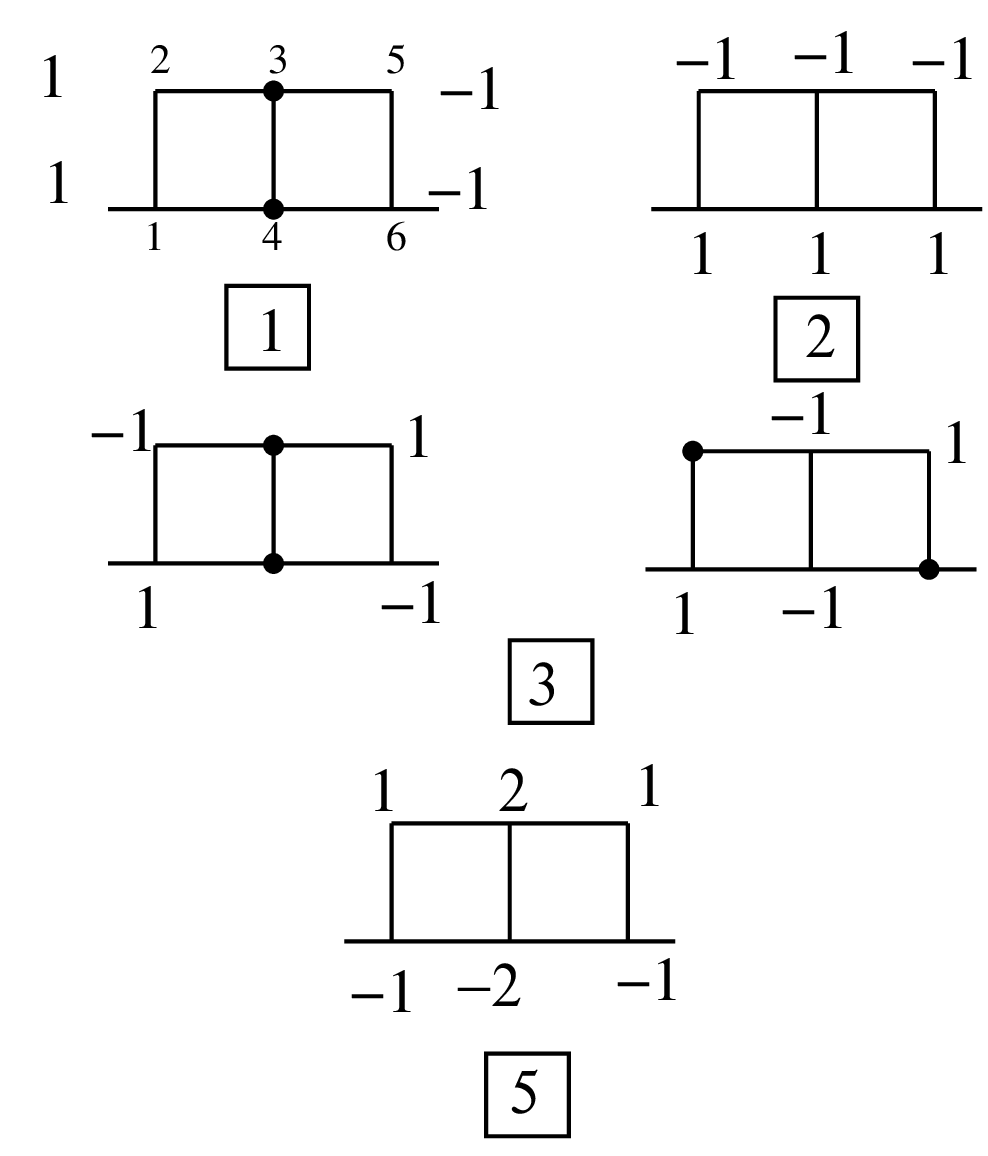,height=6 cm,width=12 cm,angle=0}
}
\caption{
Eigenvalues $\omega_i^2$ and corresponding eigenvectors $V^i$ of the graph Laplacian
for the graph shown in Fig.~\ref{gsqaa}.
}
\label{eig}
\end{figure}

The coefficients $|a|$, $|b|$, and $|c|$ as functions of $k$ are displayed in Fig.~\ref{con235}.
These results can be interpreted using the eigenvalues and eigenvectors of 
the graph Laplacian,
shown in Fig.~\ref{eig}. The eigenvalues $\omega_i^2$ are indicated in the boxes, and the components of each eigenvector $V^i_j$ are 
labeled numerically; black dots denote zero values. These components can
These components can be $1,~-1,~2,~-2$ or $0$.
For $k^2 = 2$, a lead connected at vertex 2 yields $a,b$ small and $c$
large while a lead connected to vertex 5 yields $a,c$ small and $b$ large.
The connection with the eigenvector for $k^2=2$ is not so clear here.\\
The eigenvalue $k^2 = 3$ appears in the spectral parameters for
leads connected to vertices 2 and 5. The corresponding solution
for a lead connected to vertex 2 
is the eigenvector $V$ shown to the right in the second row in Fig.~\ref{eig}. 
There $V_2=V_6=0$ so that $a=c=0$ and $b=1$. For a lead connected
to vertex 5, the solution is the same eigenvector, then $V_1=V_5=1$
so that $a=b=0$ and $c=1$. A lead connected to vertex 3 yields $b=c=0$
and $a \approx 1$ corresponding to the eigenvector shown 
to the left in the second row in Fig.~\ref{eig}.\\
The eigenvalue $k^2=5$ cannot be reached with this procedure because
$k^2 \leq 2$.

\subsection{Influence of the impedance $v$}

An initial observation from the system of equations (\ref{scatSys}) is that the coefficients $a$, $b$, and $c$ for $v \neq 1$ cannot be obtained by a simple rescaling of the results for $v = 1$. 
To see this
define ${\tilde W}=W/v,~{\tilde k}=k/v, {\tilde {\vec a}}={\vec a}/v~,
{\tilde {\vec b}}={\vec b}/v$.
The system (\ref{scatSys}) can be written as
\begin{align}\label{scatSys2}
(L_0- {\tilde W} {\tilde W}^T + {\tilde k}^2 I_n) \phi +  
{\tilde W}{\tilde {\vec b}} = - {\tilde W} {\tilde {\vec a}}  , \\
{\tilde W}^T \phi - e^{-i \alpha} {\tilde {\vec b}} = e^{i \alpha} {\tilde {\vec a}}
\end{align}
This system is the same as (\ref{scatSys}) except that the phase factor
$e^{i \alpha (k)}$ needs to be changed into $e^{i \alpha (v {\tilde k})}$ 
with $0 < {\tilde k} < 2/v$.

\begin{figure} [H]
\centerline{
   \epsfig{file=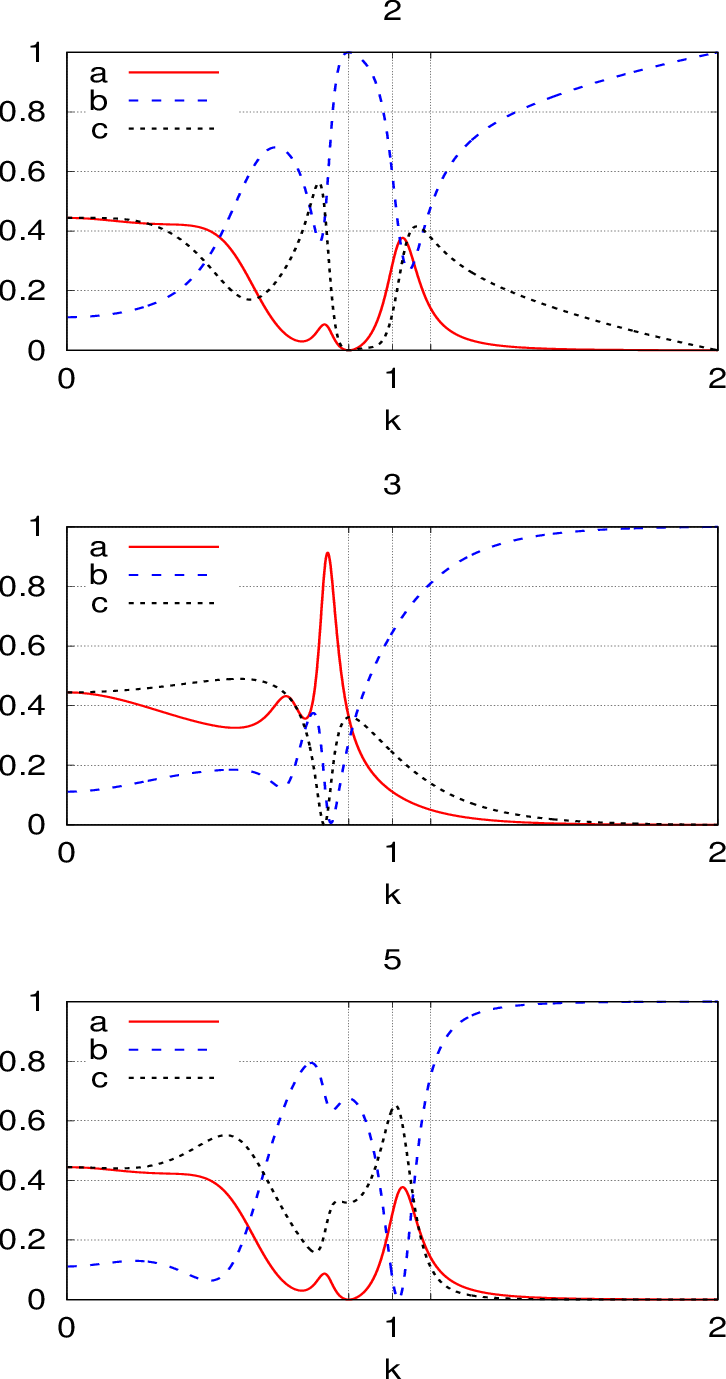,height=8 cm,width=14 cm,angle=0}
}
\caption{
Plots of $|a|^2(k)$, $|b|^2(k)$, and $|c|^2(k)$ for the graph in
Fig. \ref{gsqaa},
with the top branch is connected to vertex 2 (left panel), 3 (middle panel)
and 5 (right panel) for $v=0.5$. 
}
\label{w025con235}
\end{figure}

Taking $v=0.5$ instead of $v=1$, we obtain the results for $|a|^2, |b|^2, |c|^2$
shown in Fig. \ref{w025con235}.
The unmarked $x$-axis ticks correspond to $\sqrt{3}/2 \approx 0.866$
and $\sqrt{5}/2 \approx 1.118 $.
\begin{figure} [H]
\centerline{
   \epsfig{file=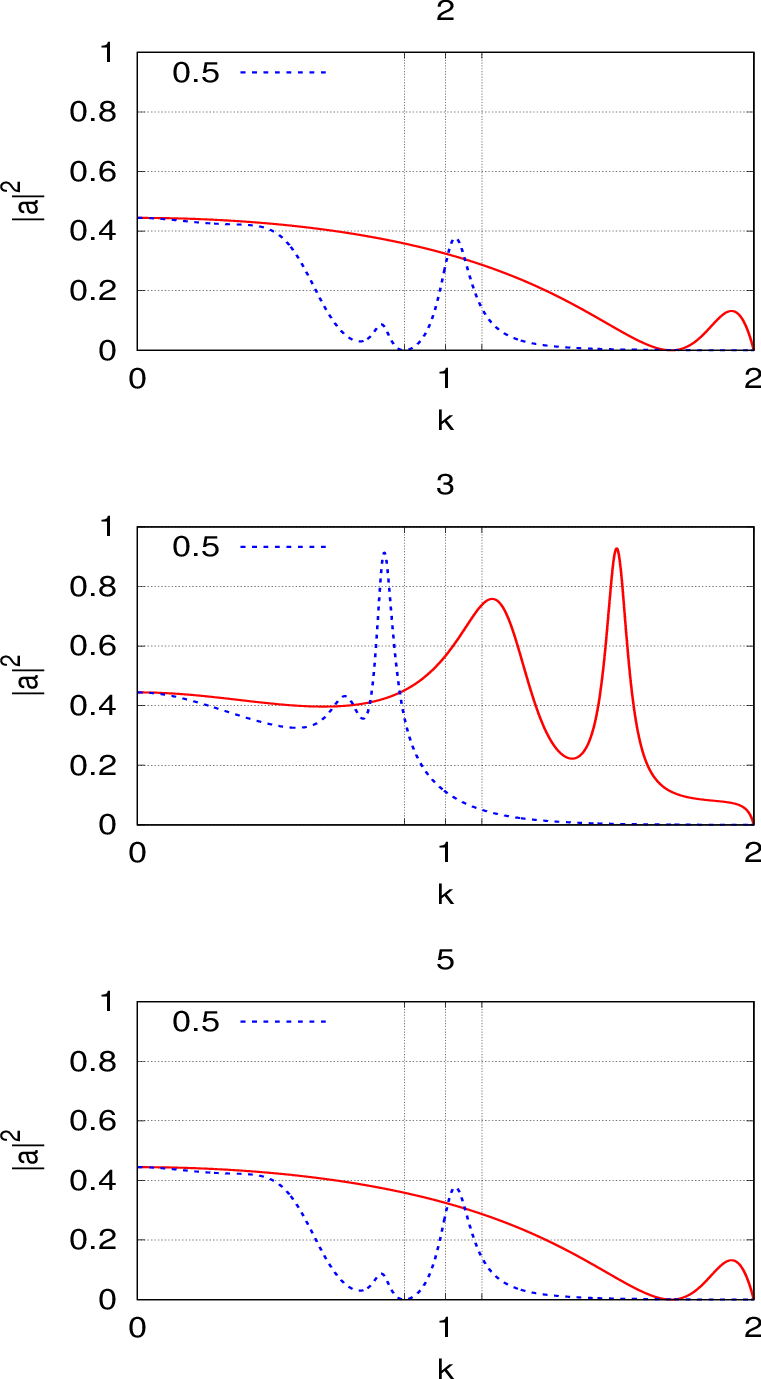,height=10 cm,width=12 cm,angle=0}
}
\caption{Plot of $|a|^2$ versus $k$ for a top branch connected 
to vertex 2 (top panel), 3 (middle panel)
and 5 (bottom panel) for $v=0.5$ and 1 (solid line).}
\label{c1c05w025con235}
\end{figure}
The unmarked ticks correspond to 
${\sqrt(3) \over 2}\approx 0.866, ~~  
{\sqrt(5) \over 2}\approx 1.118 $.

We can make the following observations based on Fig. \ref{c1c05w025con235}:\\
- As expected, leads connected to vertices 2 and 5 yield identical values for $a$ when comparing $v = 1$ and $v = 0.5$.\\
- For these two vertices, the resonance at $k = \sqrt{3}$ for $v = 1$ corresponds to a resonance at $k = \sqrt{3}/2$ (indicated by the unmarked tick left of 1) for $v = 0.5$.\\
- For a lead connected to vertex 3, $v = 1$ results in a minimum at $k = \sqrt{2}$, while for $v = 0.5$, the corresponding minimum of $|a|$ occurs at $k = \sqrt{2}/2$.

We now examine all coefficients $|a|^2(k)$, $|b|^2(k)$, and $|c|^2(k)$ for $v = 0.5$, as shown in Fig. \ref{w025con235}.

For a lead attached to vertex 2 (top panel), we observe total reflection at $k = \sqrt{3}/2$, which corresponds to an eigenvalue of $L_0$. As expected, $|a|^2(k)$ is the same for leads placed at vertices 2 and 5. 

Notably, at $k = \sqrt{2}/2$, $|c| \gg |a|$, whereas the reverse holds around $k \approx 1$. For a lead connected at vertex 5, both $|c|$ and $|a|$ are small at $k = \sqrt{2}/2$ and large near $k \approx 1$, where $|b| = 0$.

\subsection{Optimization}

Since the problem is linear, the transmission and reflection of a
wavepacket is just the integral of the transmission and reflection 
coefficients over the wave number $k$. Then, for a given input lead
we can evaluate 
the transmission of a wave packet as a function of the output lead.
\begin{figure} [H]
\centerline{
   \epsfig{file=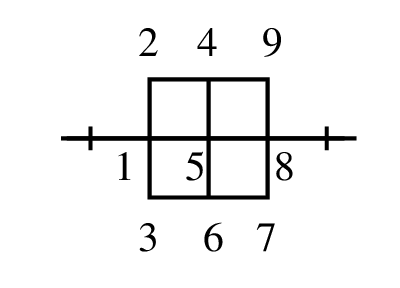,height=4 cm,width=12 cm,angle=0}
}
\caption{
A grid graph.
}
\label{g3}
\end{figure}
Consider the graph shown in Fig. \ref{g3}, we will assume the
input lead connected to vertex 1 and change the connection of the outgoing chain 
from vertices 2 to 9.
We examine the total transmission for each vertex $j$ defined as 
\be\label{ttrans} T_j = \int_0^2 |a(k)|^2 dk.\ee
The results are shown in Table \ref{tab1}. 
As expected we observe the symmetry 2-3, 4-6,7-9 so that we only 
present $T_j$ for $j=2,4,5,8$ and 9.
\begin{table}[H]
\begin{center}
\begin{tabular}{|l|r|}
\hline
$j$     &  $T_j$  \\
        &         \\ \hline
   8    &  $0.78$ \\
   9    &  $0.70$  \\
   2    &  $0.62 $ \\
   4    &  $0.47 $ \\
   5    &  $0.35$ \\
\hline
\end{tabular}
\end{center}
\caption{Total transmission for different output vertices $j$ in 
decreasing values.}
\label{tab1}
\end{table}
\noindent Note that the maximum transmission is for vertex 
8 and the minimum for vertex 5.
The value $T_8$ is almost the double of $T_5$ so that
we expect a wavepacket to be well transmitted when $j=8$ and strongly 
attenuated for $j=5$. This shows that the total transmission
of energy through a graph depends on the input and output leads.

\begin{figure}[H]
\centerline{
   \epsfig{file=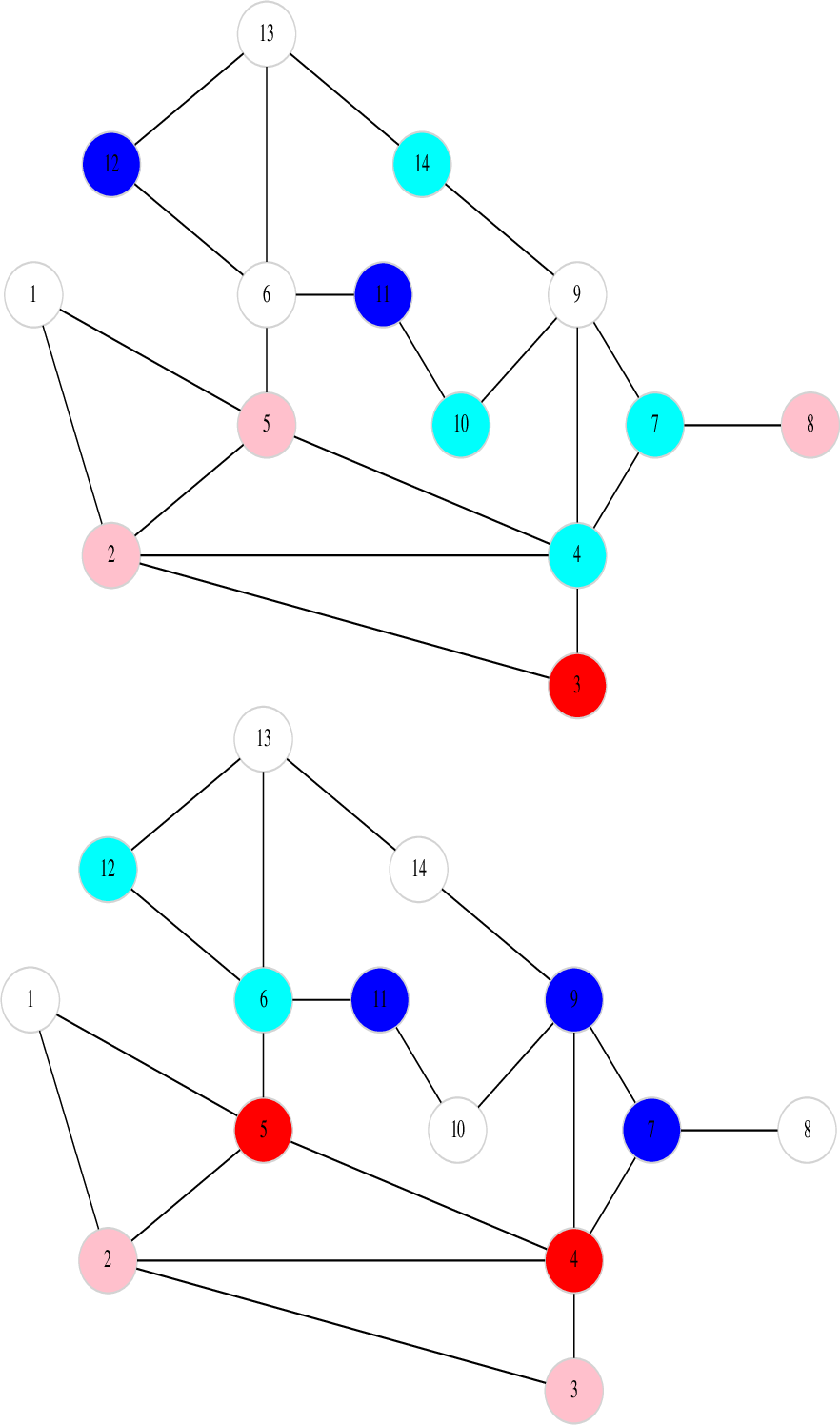, height= 7 cm, width = 12 cm, angle=0}
}
\caption{Color representation of $T_j$ for all the vertices $j$
of the graph g14 for $w=1$ (top) and $w=1/4$ (bottom). 
The color codes go from red to white.}
\label{w1w06G14}
\end{figure}

\begin{figure}[H]
\centerline{
   \epsfig{file=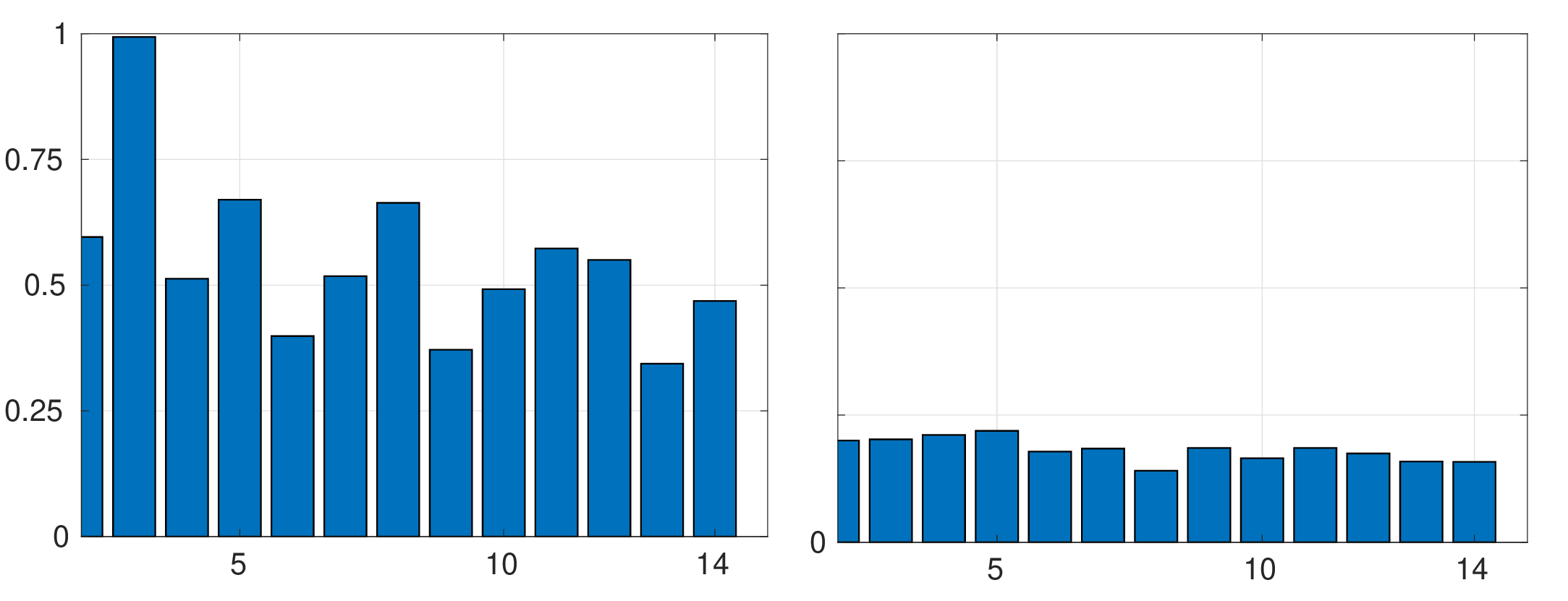, height= 7 cm, width = 12 cm, angle=0}
}
\caption{Histograms of $T_j$ for all the vertices $j$
of the graph shown in Fig. \ref{w1w06G14} for $v=1$ (top) and $v=1/4$ (bottom). }
\label{hw1w06G14}
\end{figure}

To illustrate the total transmission on a more complex example,
we consider the graph with 14 vertices shown in Fig.  \ref{w1w06G14}.
We assume there is a lead at vertex 1 and the outgoing lead is placed
at vertex $j =2, \dots, 14$.
Fig. \ref{w1w06G14} shows a color code of the total transmission
$T_j$ for the different vertices $j$. 
The histograms of $T_j$ are presented in Fig. \ref{hw1w06G14}
for $v=1$ (right) and $v=1/4$ (left). Note how the maximum transmission
for $v=1$ is almost 50 \% ($j=3$) while it is only 20 \% for $v=1/4$.
Also the transmission for $v=1/4$ varies little from one vertex $j$ to another.

\section{Detection of a Defect}

We now turn to the problem of detecting a source emitting on 
a graph and demonstrate how scattering can provide useful insights.
This corresponds to a non-zero term $F_k$ in equation (\ref{ghelm}); specifically, we choose
\be \label{defect}  F_k = d \exp(-w_d (k-k_d)^2). \ee
We again consider the graph shown in Fig. \ref{g3}, placing a defect at a vertex $i_d$ at a given frequency $k_0$.
We then compute the scattering parameters $a$ and $b$. While it is difficult to extract meaningful information directly from these parameters, the quantity $|a|^2 + |b|^2$ proves to be more informative. 
We observe that this quantity is no longer constant and equal to one; instead, it varies in the region around $k_0$.


\begin{figure} [H]
\centerline{
   \epsfig{file=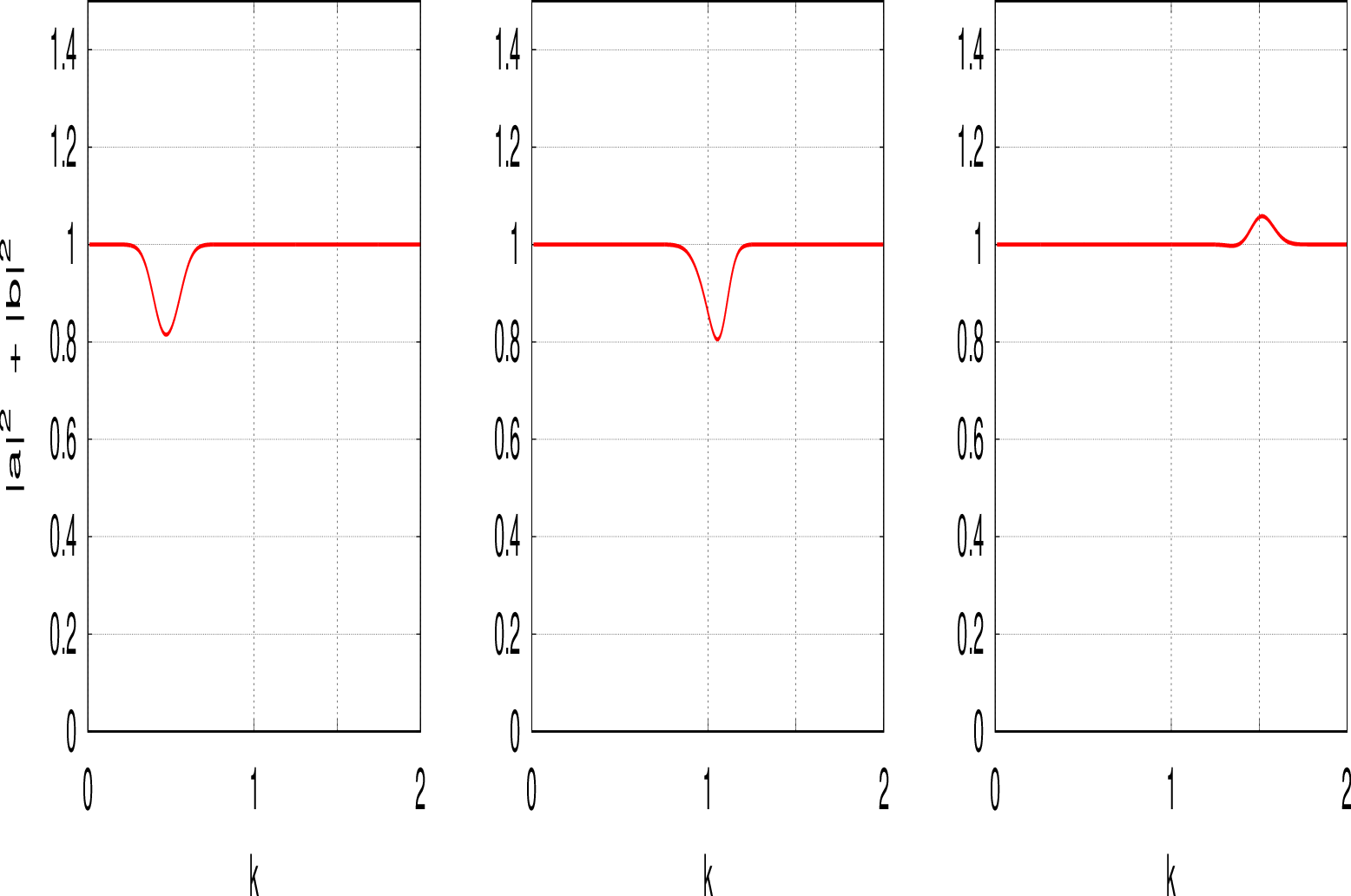,height=12 cm,width=5 cm,angle=90}
}
\caption{Plot of $|a|^2+|b|^2$ for a defect placed at positions
2, centered at $k_d=0.5,1$ and $1.5$ from left to right.}
\label{dk}
\end{figure}

Fig. \ref{dk} shows $|a|^2+|b|^2$ as a function of $k$ for a defect placed at vertex $i_d=2$
in the graph shown in Fig. \ref{g3}, for $k_d=0.5,1$ and $1.5$.
As expected $|a|^2+|b|^2$ deviates from 1 in a region centered on $k_d$.
This region is not necessary symmetric as shown in the plots.
Moreover, the deviation can be either positive or negative.

\begin{figure} [H]
\centerline{
 \epsfig{file=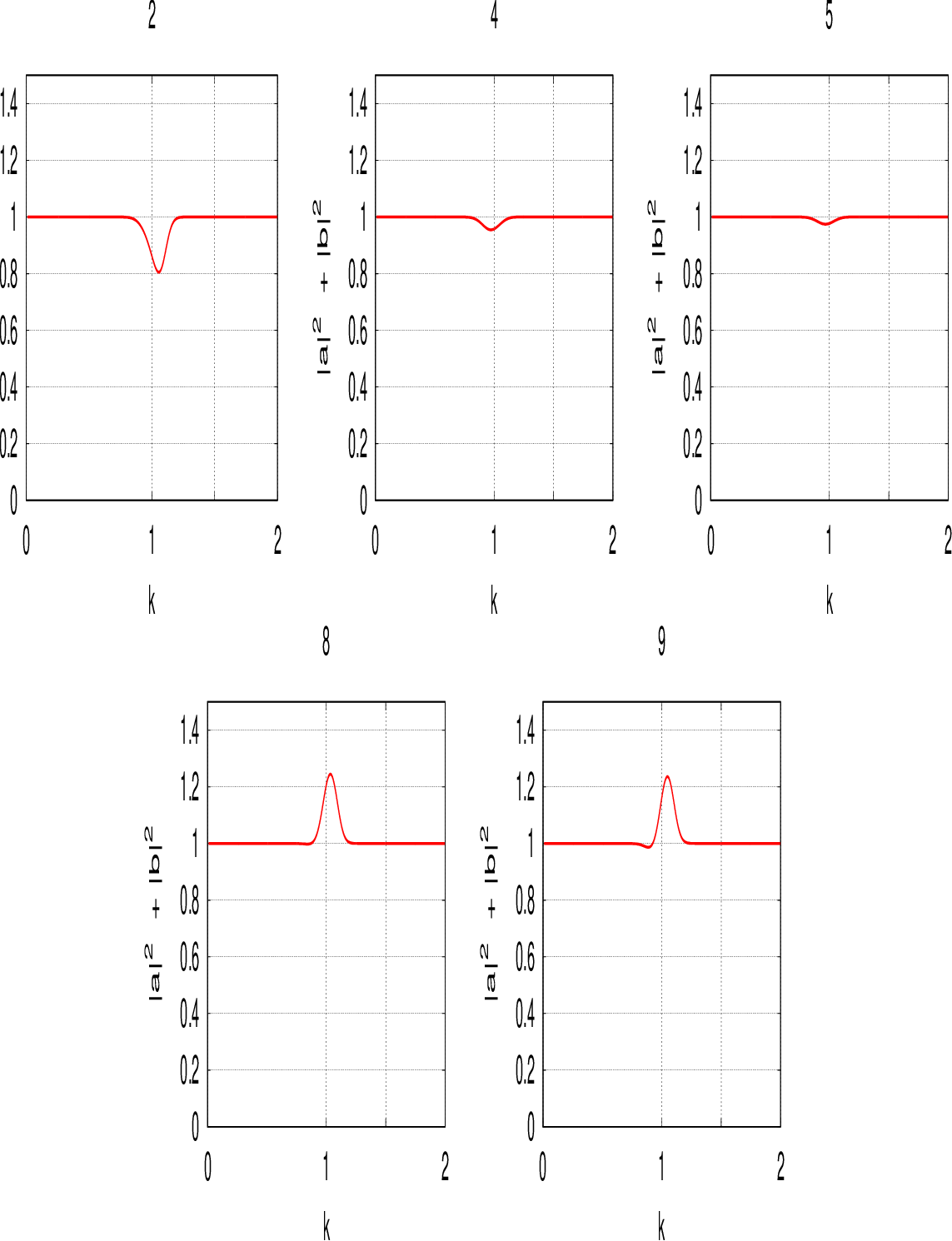,height=8 cm,width=12 cm,angle=0}
}
\caption{Plot of $|a|^2+|b|^2$ for a defect placed at positions
2,4,9,5 and 8 from top left to bottom right.}
\label{dpos}
\end{figure}

Keeping $k_d=1$ fixed, we examined the influence of the lead vertex $i_d$.
The results are shown in Fig.  \ref{dpos}. As expected, the 
variation of $|a|^2+|b|^2$ is concentrated around $k_d=1$.  Its shape depends on the choice of $i_d$, suggesting that, in principle, one could determine which vertex is affected by the defect.

\section{Conclusion}

This article investigates wave scattering by a combinatorial graph connected to multiple external leads. The scattering formalism is simplified, and a linear algebra-based algorithm is introduced to determine reflection and transmission coefficients as a function of wavenumber, $k$. This method is initially applied to the graph Laplacian but can be readily generalized to the adjacency matrix.

Numerical analysis of the scattering reveals the influence of different lead vertices on the spectral reflection and transmission properties. Notably, total reflection arises from the presence of bound states, identifiable as eigenvectors of the graph Laplacian. Reducing the lead impedance shifts the reflection and transmission coefficients toward smaller $k$ values, specifically displacing a resonance at $k_{0}$ for unit impedance to $v k_{0}$. Furthermore, specific lead configurations are shown to maximize total transmission across the entire range  $0\le k\le 2$.

Finally, we address the practical inverse problem of detecting a localized source  on a graph using this scattering framework. The resulting reflection and transmission spectra encode information about both the spectral content and the vertex location of the defect.

\end{document}